# Apollonius Solutions in $\mathbf{R}^d$
## Power Diagram Connections

Raymond P. Scaringe

## Abstract

Voronoi and related diagrams have technological applications, for example, in motion planning and surface reconstruction, and also find significant use in materials science, molecular biology, and crystallography. Apollonius diagrams arguably provide the most natural division of space for many materials and technology problems, but compared to Voronoi and power diagrams, their use has been limited, presumably by the complexity of their calculation. In this work, we report explicit equations for the vertices of the Apollonius diagram in a d-dimensional Euclidean space. We show that there are special lines that contain vertices of more than one type of diagram and this property can be exploited to develop simple vertex expressions for the Apollonius diagram. Finding the Apollonius vertices is not significantly more difficult or expensive than computing those of the power diagram and have application beyond their use in calculating the diagram. The expressions reported here lend themselves to the use of standard vector and matrix libraries and the stability and precision their use implies. They can also be used in algorithms with multi-precision numeric types and those adhering to the exact algorithms paradigm. The results have been coded in C++ for the 2-d and 3-d cases and an example of their use in characterizing the shape of a void in a molecular crystal is given.

## Introduction

The Voronoi diagram is a partition of space used extensively in scientific and technological applications. A review by Aurenhammer and Klein [AK00] provides an accessible introduction to the subject and has been updated [AKL13] in book form. The book by Okabe *et al.* [OBSC00] is also a standard source and includes applications to data analysis. As emphasized in the above monographs, while the Voronoi diagram proper is a partitioning based on a point set, the concept can be generalized to sets of other geometric objects, of particular relevance here, to circles (2-d), spheres (3-d) and so on, often simply referred to as balls. Two different diagrams resulting from the generalization from points to balls are in common use and each has a variety of names [FN1]. For the first type we follow Aurenhammer [Aur87] in using the term power diagram and for the second, Emiris and Karavelas [EK06], and use the term Apollonius diagram. The work reported here focuses on the development of exact algebraic solutions but the treatment is in no way limited to the exact computation paradigm; both the notation and discussions support algorithms ranging from ordinary floating point to exact computational designs. Although a variety of applications are discussed below, explanations and examples are somewhat biased toward atomic scale applications. This is not meant to imply that other applications are in some way less important, but rather reflects the interests of the author.

At atomic scale, replacing a point by a ball means augmenting an atom position with an atomic radius and this idea has found extensive use in solid state chemistry, molecular biology, and materials science, not only in the form of the space-filling diagram, but also in the form of power and Apollonius diagrams. A relatively early example by Fisher and Koch [FK79] uses power diagrams to study the structure types exhibited by molecular crystals. They have also been used in the calculation of atomic volumes in molecular crystals [KF80] and in proteins [GF82]. Power diagrams and their duals also underly the concept of weighted alpha shapes [Edel95], which are especially well suited to characterizing a union of balls, for example, the space filling diagrams of molecules as large as proteins [LEFSS98, LEW98]. Apollonius diagrams have been used in the study of large scale structures, particularly, liquids/amorphous solids [AAMVJ04, AVMMPJ05, VBS02], and proteins [GPF97, Willl98, KKS02]. Many of these studies focus on the characterization of voids, channels, binding pockets, and interfaces. Applications to proteins have been reviewed by Poupon [Pou04]; some materials applications appear in the book edited by Gavrilova [Gav08] and others are cited in [MVLG06].

The distinction between the various diagrams is illustrated by 2-d examples in Figures 1 and 2 [FN2]. In the case of a point set on a plane, the Voronoi diagram (Figure 1a) consists of cells (also called regions), each of which is a convex polygon (in 3-d convex polyhedron) and each of which is associated with exactly one site. Each cell encloses all the points in the plane that are closer to one site than to any of the other sites in the set. It is then clear how the cell geometry can be used to calculate the area occupied by each site in the plane (or in 3-d, volume occupied) and provides a natural definition for the term neighbor, that is, two sites that share an edge (in 3-d, face). For a set of sites that are circles, the power diagram (Figure 1b) and Apollonius diagram (Figure 1c) still partition the plane into cells, one per site, but not in the same way. In the power and Apollonius diagrams, sites represented by larger circles tend to occupy larger cells than in the corresponding Voronoi diagram. Here, simply note that like the Voronoi cell, the edges of the power cell are straight lines and vertices occur at line intersections. In the case of the Apollonius diagram, the edges can be curved and the vertices occur at the intersection of these curves.

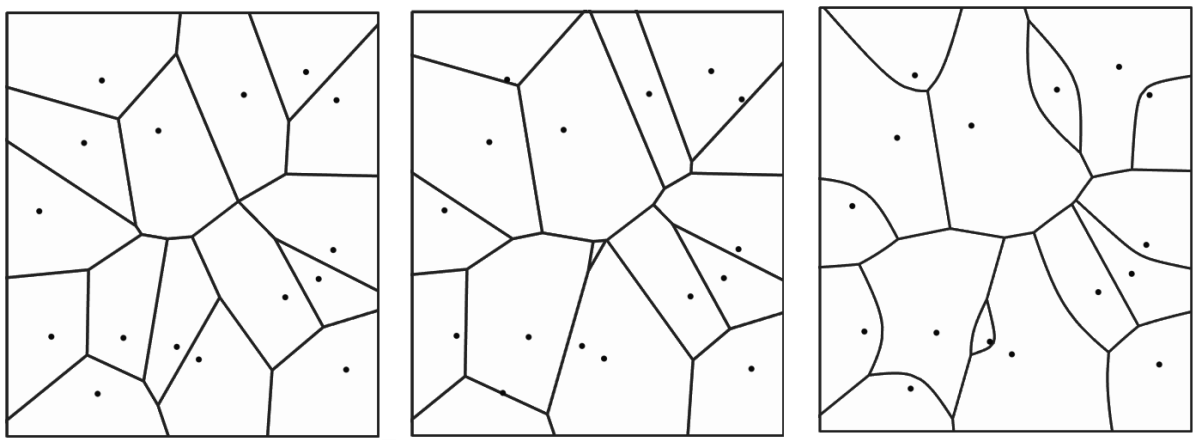

*Figure 1 Diagrams: site centers – points, edges – lines; (a-left) Voronoi, (b-center) Power, (c-right) Apollonius*

Each vertex in the diagrams of Figure 1 is determined by a distance metric that defines the term "closer to" (see Sections I and II). This same metric can be used to define a radius for each vertex, so that each vertex is also a ball; the sites can then be called generators for a clear distinction. For 2-d, each vertex in the diagram is "closer to" to at least three balls of the generator set than to any of the other generators. In 3-d, the corresponding condition is closer to at least four balls, and so on in higher dimensions. This condition implies a specific geometric relationship (Figure 2) between the vertex ball and the generator balls "closest" to it; in particular, they are each equidistant to the vertex based on the distance metric at hand. These properties are central to construction of the diagrams but are also used directly in various applications [AKL13, OBS00]. For example, given only a point sampling of an object surface, the Voronoi balls (Figure 2a) of the point set can be used to reconstruct the entire surface and also approximate the medial axis of the object itself [AmK00]. The power balls (Figure 2b) can be used in defining the "dual diagram" which in turn can be used to create surface tessellations from point samplings [ACK01]. Weighted alpha shapes can be used to construct the medial axis of the same object [AK01] and a complex similar to the weighted alpha shape can be used to determine the λ-medial axis [ABE09]. We add here that all of the properties that make these applications possible are not obvious from a visual examination of the Voronoi or power balls overlaid on the diagrams as in Figures 2a-b.

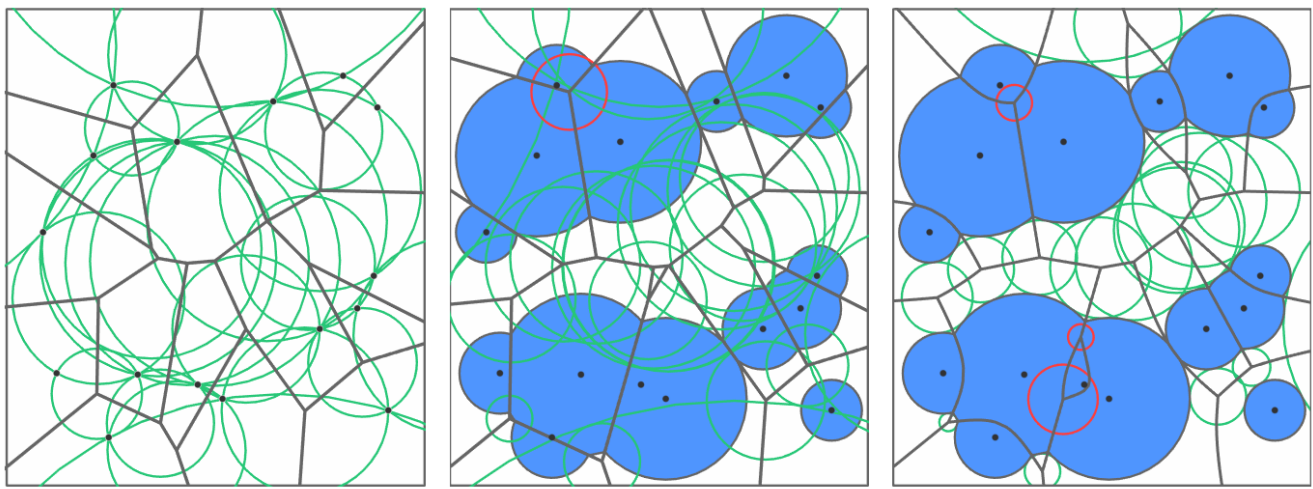

*Figure 2 Generator (blue) and Vertex (green/red) Balls Overlaid on the Diagrams of Fig. 1:*
*(a-left) Voronoi, (b-center) Power, (c-right) Apollonius*

The Apollonius balls are shown in Figure 2c. Here, the interest in the field of motion planning (see [AKL13], pp. 216-222) is not hard to grasp. One simply imagines the need to navigate a circular vessel, left to right, through the collection of circular obstacles in Figure 2c. One finds the path of maximum clearance is along the cell edges that separate the top part of the ball collection from the bottom; each vertex along this path marks a point requiring a course correction. The connection to realistic applications follows from the fact that the shape of any solid object can be recreated with a sufficiently large collection of balls [ACK01, GMP07]. In the context of materials, molecular shapes are almost always represented by a collection of balls that represent atomic positions and intermolecular radii. The edges of the diagram then represent the trajectories of least resistance to mass transport. Also note that the Apollonius balls emulate the principles of close packing, in every detail, as first described by Kitaigorodsky [Kit73, Kit61]. Hence, a void filled with Apollonius balls is a kind of properly oriented superposition of the space filling models of molecules that could occupy the void. It should also be possible to use Apollonius solutions to calculate the medial axis of plane curves bounded by circular arcs which have been investigated for this purpose [AAAHJR09, AAHW11]. Moreover, the medial axis of a void formed by a union of balls is completely contained in the Apollonius diagram.

Aurenhammer *et al.* [AKL13] point out that the Apollonius diagrams have received only moderate attention. In fact, the methods used to calculate the Apollonius diagram often rely on the construction of a different diagram altogether. Boissonnat *et al.* describe its construction using the convex hull [BK03, BD05]. Other methods include the intersection of cones [Aur87] with the lifted power diagram and lower envelope calculations [Will98, Will99, HE09]. Aurenhammer's lifting method has been implemented

[ABMY02] in 2-d. Karavelas and Yvinec [KY06], using the predicates developed in [EK06], derive the 2-d Apollonius diagram from its dual. It is also reported [EK06] that the Apollonius diagram can be obtained as a concrete case of the abstract Voronoi diagrams of Klein *et al.* [KMM93].

The edges of the diagram have been characterized in a number of ways. For the 3-d case, Will [Will99] reported a detailed study of the analytical properties of the Apollonius edges; these edges are projected onto a unit sphere from which a single cell of the diagram is derived. Hanniel and Elber [HE09] develop the edge as a ratio of rational polynomials and these are projected onto a unit cube, again resulting in a single cell of the diagram. In 3-d, Kim *et al.* [KCK05] have developed a rational spline representation of 3-d edges which are pushed onto an "edge stack". In that work and in the algorithm of Medvedev *et al.* [MVLG06], diagram construction proceeds by edge (channel) tracing to find site sets suitable for vertex analysis.

Vertices seem to have received rather less attention. In 2-d, the idea of finding the intersection of two hyperbolas in the plane has been investigated by applying a symbolic algebra program to the problem; Kim *et al.* [KKS02] report that the resulting equations require a 3MB ASCII file to contain them and result in 70,000 lines of C code. In the same paper they derive expressions by Möbius transformations which individually are not complicated. However, six different configurations are distinguished for the case of all unequal generator radii, five for the case of two generator radii equal, and also special cases for generators of zero radius. With a different approach [ABMY02] to the same problem, the final solution equations require only about one journal page. Emiris and Karavelas [EK06] also report compact 2-d expressions using the method of inversion. Not surprisingly, vertex solution methods for the case of spheres in 3-d [AMS11] and also higher dimensional cases [Gav09] are rare and more complicated than those for the corresponding Voronoi and power diagrams (see Section II). Whether or not an inversion technique is used, vertex radii result from the solution of a quadratic equation. However, we find little discussion of the connection between roots and the geometry of ball configurations even though this connection has important implications for the diagram. Similarly, we find little discussion of special cases or their root characteristics.

Apparently, the approach taken in determining the vertices of the Apollonius diagram can have a significant effect on the complexity of the solution. Solution complexity affects the time required to design an appropriate algorithm, to write code, to analyze special cases, to assess numerical stability, and will very likely contribute to longer run times. It is an obvious deterrent to routine application of Apollonius diagrams to problems where the use of Voronoi and power diagrams have become quite routine. For example, Gavrilova and Rokne [GR03] mention that use of Apollonius diagrams were dismissed on the basis of complexity for a motion planning application. In cases where Apollonius diagrams are considered indispensable, algorithms have been devised [MVLG06, KCK06] that make efficient use of hyperbolic paths to find candidate ball sets, but this does not eliminate the need for vertex calculations. For example, Medvedev *et al.* [MVLG06] report that with their 3-d implementation, which runs in O(n), most of the computational effort is still spent in vertex calculations. Vertex positions and radii are always required, in one form or another, for construction of the Apollonius diagrams.

Here, we examine the properties of Apollonius solutions as viewed from the power vertices of the same ball set in the same dimension. The solutions obtained satisfy a particular tangency relationship suitable to the Apollonius diagram, but somewhat more general. By including all possible tangency relationships, the same methods can be used to solve the original $10^{th}$ problem of Apollonius and its d-dimensional generalizations [FN3]. All solutions reported here are exact in the usual algebraic sense that no approximation of any kind is used in their derivation. The report is organized as follows: Section I contains definitions, conventions, the problem description, and two solution methods using lifted vectors. Section II examines the connection between the Apollonius vertices of a ball set and power vertices of the same and related ball sets, and describes a third method. In Section III, various details important for implementation are discussed, "Recipes" are given, and an example of using Apollonius vertices to investigate the voids of a molecular crystal is given. An overview of main results is given in Section IV.

# I. Problem Description in $\mathbf{R}^d$ and Solutions using Lifted Vectors in $\mathbf{R}^{d+1}$

## A. Definitions and Notation

The notation used throughout is conventional: reference to a point, q, in a d-dimensional space, $\mathbf{R}^d$, implies coordinates, $(q_1, q_2, ..., q_d)$, associated with the position vector, $\mathbf{q} = q_1\hat{\mathbf{e}}_1 + q_2\hat{\mathbf{e}}_2 +...+ q_d\hat{\mathbf{e}}_d$, where $\hat{\mathbf{e}}_i$ is the unit vector parallel to the ith coordinate axis of an orthonormal system, and $|\mathbf{q}|$ is the norm of $\mathbf{q}$ with $|\mathbf{q}|^2 = |\mathbf{q}\cdot\mathbf{q}| = q_1^2 + q_2^2 +...+ q_d^2$. If a point is one of a set, and therefore its position vector subscripted, $\mathbf{q}_i$, the coordinates are denoted $(q_{i1}, q_{i2},...q_{id})$. Square matrices are indicated with bold underlined upper case letters, *e.g.* $\underline{\mathbf{A}}$, and column matrices as bold underlined lower case letters, *e.g.* $\underline{\mathbf{c}}$.

Let **B** be a set of balls in $\mathbf{R}^d$, $b_i = (x_i, r_i)$, $1 \le i \le (d + 1)$, $x_i \in \mathbf{R}^d$ and $r_i \in \mathbf{R}$. The (minimum) distance, $d_a$, between the surface of the ball, $b_i$, and a point, $x \in \mathbf{R}^d$, is given by:

$$d_a(b_i, x) = |\mathbf{x}_i - \mathbf{x}| - r_i$$

A ball, $b = (x, r)$, $x \in \mathbf{R}^d$, $r \in \mathbf{R}$, that is tangent to $b_i$, then satisfies the equation,

$$|\mathbf{x}_i - \mathbf{x}|^2 = (r_i + r)^2 \qquad (1)$$

or upon expansion and rearrangement,

$$2\mathbf{x}_i\cdot\mathbf{x} + 2r_i r = |\mathbf{x}_i|^2 - r_i^2 + |\mathbf{x}|^2 - r^2 \qquad (2)$$

If, b, is mutually tangent to every ball in **B**, it satisfies the d + 1 simultaneous equations (2) with $1 \le i \le (d + 1)$; such a ball is referred to as a solution ball of **B**. As is obvious from Figure 1, if the d + 1 balls of **B** are chosen randomly from the entire generator set, there is a high probability that the solution ball(s) of **B** will not be vertices of the final diagram. Nevertheless, each vertex of the final Apollonius diagram is a solution ball of some set **B**. Also note that in order to proceed to analytical solutions both sides of the tangency condition have been squared so there are normally two solution balls for every set **B**. In many problems this would mean that one solution is always spurious, but for Apollonius diagrams both solutions can be vertices of **B**, and both can also occur in the final diagram. Moreover, the occurrence of two solutions can be used to advantage in devising solution methods (see Section I.D).

In some treatments it is not allowed that the balls of **B** overlap [AMG98], [HE09]. Here this is allowed because, for example, the use of a union of medial balls representing a sampled solid object typically result in many closely overlapping balls. Similarly, the van der Waals spheres of covalently bonded atoms always overlap (see also Section III.C) Many treatments require that no ball in **B** can be completely contained by another, but this restriction is also unnecessary here (see Section III). Since such balls have no cell in the diagram they are often termed "trivial" in the literature. Also, a ball can be "hidden" in the sense that it is completely contained in the union of two or more of the other balls of **B** but not trivial (see for example, the smaller of the three balls in Figure 3f). However, it is required that no two balls be identical or concentric. Otherwise, no restrictions are necessary to proceed from (2) to vertex solutions.

As is well known, incrementing radii of all generator balls by the same quantity does not change the Apollonius diagram (Figure 1c), so here, all generator radii are assumed positive or zero. That this can be done without loss of generality can be understood as follows: suppose the ball $b = (x, r)$ is a solution to (1) for some ball set **B** and we then define a second ball set, **B'**, by adding the same radial increment, $r_\varepsilon$, which can be positive or negative, to each of the balls of **B** but without changing their position vectors: $b'_i = (x_i, r_i + r_\varepsilon)$. Then, $b' = (x, r - r_\varepsilon)$ is a solution to (1) for the set **B'**, because none of the vectors on the left hand side have changed nor has the sum on the right hand side. For example, Figure 1c was calculated using the ball radii as shown in Figure 2c. If we were to increment each generator ball depicted in Figure 2c by the same arbitrary amount, $r_\varepsilon$, Figure 1c would not change in any way, but with $r_\varepsilon > 0$, the generators (blue disks) of Figure 2c would all be larger and the solution balls (green circles) correspondingly smaller. The requirement of non-negative generator radii is a convention that simplifies the analysis and discussion of roots, and the illustration of the solution geometry. However, the solution equations are completely general and applicable to any ball set whether it contains generator balls with negative radii or not.

## B. Lifted Vector Solution

Except for the extra term, $r_i\, r$, on the left hand side, (2) is identical to the more easily solved equation for the power vertex of the ball set **B** (see Section II). Accordingly, we apply a change of origin as follows:

$$\mathbf{v}_i = \mathbf{x}_i - \mathbf{p} \tag{3a}$$
$$\mathbf{v} = \mathbf{x} - \mathbf{p} \tag{3b}$$

where **p** is the position vector of the power vertex of **B**. The extra scalar term on the left hand side of (2) can be absorbed by defining lifted vectors in $\mathbf{R}^{(d+1)}$ as follows:

$$\boldsymbol{\alpha}_i = \mathbf{v}_i + r_i \hat{\mathbf{e}}_{d+1} \tag{4a}$$
$$\boldsymbol{\alpha} = \mathbf{v} + r \hat{\mathbf{e}}_{d+1} \tag{4b}$$

where $\hat{\mathbf{e}}_{d+1}$ is the unit vector along the lifted axis so that $\hat{\mathbf{e}}_n \cdot \hat{\mathbf{e}}_{d+1} = 0$, $1 \leq n \leq d$. Here and throughout, all lifted vectors are denoted by bold Greek letters. Note that these lifted vectors have the same form as those used in [ABMY02] in the 2-d implementation of Aurenhammer's algorithm [Aur87]. Here, all the radius information about the d-balls is considered to be encoded in the last component of the lifted vector so unlike the above methods, we do not define lifted radii in order to define (d + 1)-balls in $\mathbf{R}^{d+1}$. With (3) and (4), (2) is rewritten as follows:

$$2\boldsymbol{\alpha}_i \cdot \boldsymbol{\alpha} = |\mathbf{v}_i|^2 - r_i^2 + |\mathbf{v}|^2 - r^2 \tag{5}$$

where $|\mathbf{v}_i|^2 - r_i^2$ is the power distance of $b_i$ from the origin (Section II). The quantity, $|\mathbf{v}|^2 - r^2$, has the analogous meaning for the solution ball(s) and contains quadratic terms in all unknown quantities. These terms can be temporarily eliminated by subtracting the equation for the kth ball from (5), where any ball can be taken as the kth providing $i \neq k$, and the result is as follows:

$$2\boldsymbol{\alpha}_{ik} \cdot \boldsymbol{\alpha} = (|\mathbf{v}_i|^2 - r_i^2) - (|\mathbf{v}_k|^2 - r_k^2), \text{ with: } \boldsymbol{\alpha}_{ik} = \boldsymbol{\alpha}_i - \boldsymbol{\alpha}_k \tag{6}$$

There are, at most, d linearly independent equations like (6) as compared to the d + 1 components of **α** that are unknown, in other words, insufficient for direct solution. As regards origin choices in $\mathbf{R}^d$, one would arrive at an equation like (6) without making any origin change, or, by making any origin change other than that made in (3). However, with the power vertex, **p**, as the origin, the power distances must satisfy an additional condition, namely: $|\mathbf{v_i}|^2 - r_i^2 = r_p^2$, $1 \leq i \leq d + 1$, where $r_p^2$ is a constant equal to the square of the power radius defined by the set **B** (see Section II, below (P5)). The right hand side of (6) therefore vanishes leaving a set of d homogeneous difference equations in d + 1 unknowns.

$$\boldsymbol{\alpha_{ik}} \cdot \boldsymbol{\alpha} = 0 \tag{7}$$

The meaning of (7) is that in $\mathbf{R^{d+1}}$, the lifted solution vector $\boldsymbol{\alpha}$ is mutually perpendicular to each of the d lifted difference vectors $\boldsymbol{\alpha_{ik}}$. Evidently, any vector, $\boldsymbol{\alpha'}$, parallel to $\boldsymbol{\alpha}$, also satisfies (7). For example, for d = 2, that is, three circles in the xy plane, the three lifted vectors, $\boldsymbol{\alpha_i}$, represent a triangle whose vertices lie on a plane in a kind of 3-space where the radii serve as "z" coordinates. A vector normal to this plane can be found by taking the cross product of any two difference vectors:

$$\boldsymbol{\alpha'} = \boldsymbol{\alpha_{ik}} \wedge \boldsymbol{\alpha_{jk}}, \qquad 1 \leq i \neq j \neq k \leq 3$$

A method for finding a (d + 1)-component vector, $\boldsymbol{\alpha'}$, normal to a set of d vectors, each with d + 1 components, is given in Appendix I; it is strictly analogous to usual matrix development of the cross product for 3-component vectors.

The lifted vector, $\boldsymbol{\alpha'}$, is therefore considered known, but since it is not necessarily equal to the solution vector, $\boldsymbol{\alpha}$, but simply parallel, only its directional information is used through definition of the unit vector, $\boldsymbol{\hat{\alpha}}$, in the $\boldsymbol{\alpha'}$ direction,

$$\boldsymbol{\hat{\alpha}} = \boldsymbol{\alpha'}/|\boldsymbol{\alpha'}| \tag{8}$$

and the lifted solution vector, $\boldsymbol{\alpha}$, in terms of $\boldsymbol{\hat{\alpha}}$ is:

$$\boldsymbol{\alpha} = \sigma \boldsymbol{\hat{\alpha}} \tag{9}$$

This leaves the scale factor, $\sigma \in \mathbf{R}$, as the final unknown to be determined. From the above definitions, we note: $|\mathbf{v}|^2 = |\boldsymbol{\alpha}|^2 - r^2$, $r^2 = \sigma^2(\boldsymbol{\hat{\alpha}} \cdot \mathbf{\hat{e}_{d+1}})^2$, and $|\mathbf{v_i}|^2 - r_i^2 = r_p^2$, so that (5) can now be written in terms of $\boldsymbol{\hat{\alpha}}$ and $\sigma$,

$$\sigma^2(1 - 2(\boldsymbol{\hat{\alpha}} \cdot \mathbf{\hat{e}_{d+1}})^2) - 2\sigma \boldsymbol{\hat{\alpha}} \cdot \boldsymbol{\alpha_i} + r_p^2 = 0 \tag{10}$$

and $\sigma$ is determined from the quadratic formula:

$$\sigma_\pm = \{\boldsymbol{\hat{\alpha}} \cdot \boldsymbol{\alpha_i} \pm [(\boldsymbol{\hat{\alpha}} \cdot \boldsymbol{\alpha_i})^2 - (1 - 2(\boldsymbol{\hat{\alpha}} \cdot \mathbf{\hat{e}_{d+1}})^2)r_p^2]^{(1/2)}\}/(1 - 2(\boldsymbol{\hat{\alpha}} \cdot \mathbf{\hat{e}_{d+1}})^2) \tag{11}$$

If the discriminate in (11) is negative, there are no real solutions and the kinds of ball arrangements that give rise to this depend on the space dimension, d (see Sec. III for a discussion). Otherwise, there are two real solutions which we denote $\sigma_-$ and $\sigma_+$, and the related lifted solution vectors are:

$$\boldsymbol{\alpha_\pm} = \sigma_\pm \boldsymbol{\hat{\alpha}} = \mathbf{v_\pm} + r_\pm \mathbf{\hat{e}_{d+1}} \tag{12}$$

## C. Solution Analysis

To simplify the discussion of roots we require that the inner product $\boldsymbol{\hat{\alpha}} \cdot \mathbf{\hat{e}_{d+1}} \geq 0$. That is, if component d + 1 of **α'** in (8) is less than zero we replace equation (8) with **α̂** = - **α'**/|**α'**|. Either choice for **α̂** will clearly satisfy (7), but the above requirement ensures that the sign of the solution radius, r, is identical to the sign of σ. Then, any description of roots based on the sign of r is equivalent to one based on the sign of σ. The scalar, σ, is unique to (9 – 11) while the solution radius, r, occurs in all methods reported here. Accordingly, we analyze roots in terms of, r, so a single discussion will suffice for all methods. The notation $σ_+$, $σ_-$, $r_+$, $r_-$, etc., only indicates which of the two roots in (11) is being singled out and all such scalars can be positive or negative.

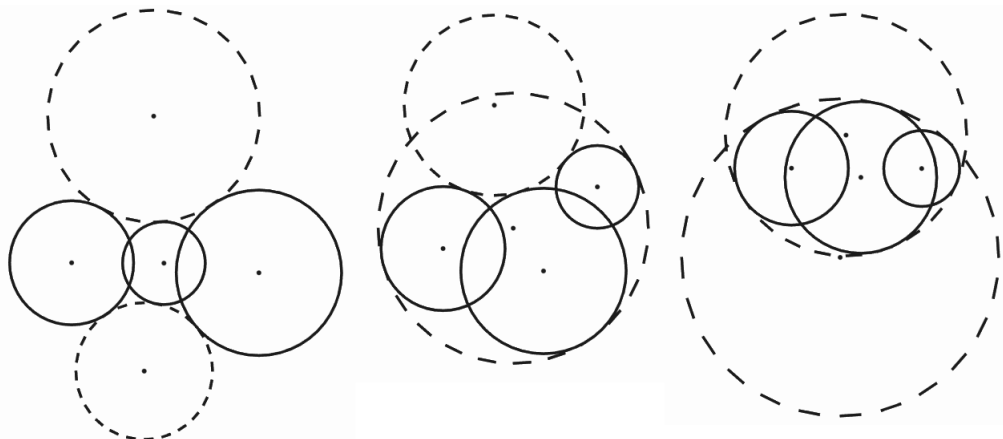

*Figure 3: Solutions-2d; generator balls are solid, solution balls are dashed*
*above: (a-left), (b-center), (c-right); margin: (d-top), (e-middle), (f-bottom)*

Also, the radii, $r_+$ and $r_-$ can be positive or negative in any combination. Solutions can be further characterized by the presence or absence of large negative solutions, that is, solution balls that envelop all the balls of **B.** There are then six combinatoric possibilities and each has a geometric realization as shown in Figure 3 for the 2-d case.

With both roots positive (Figure 3a) there are two solution balls with positive radii, both mutually tangent to all the balls in **B** and both exclude all the balls in **B** from their interior. If $r_+ > 0$, but $r_- < 0$, there is just one solution ball that excludes the balls in **B**. The second can be a large negative ball that includes all the balls of **B** in its interior (Figure 3b), or a small negative ball that is included by all of the balls of B (Figure 3d). If $r_+$ and $r_-$ are both negative there are two solution balls, both with negative radius and then there are three possibilities: Either both include all the balls in **B** (Figure 3c), or just one does and the second is included by all the balls in **B** (Figure 3e), or both are included in all of the balls of **B** (Figure 3f). Diagram examples of small negative vertices are shown as red circles in Figure 2c (also see Figure 8).

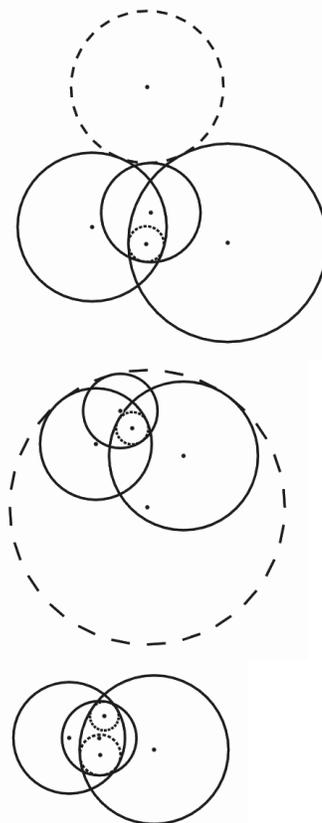

The analysis of solutions suited for the Apollonius diagram is straightforward: all solution balls shown in Figure 3 can be a valid part of the diagram except those with large negative radii. Large negative solution balls arise from the "wrong branch" of the hyperbola (2-d), hyperboloid (3-d), and so on; that is, from the branch that is not used in the definition of hyperbolic half spaces. Obvious from the above, but interesting, is that any given solution pair may or may not contain vertices relevant to the diagram problem. At this point it is also apparent that the sign of a solution encodes a mode of tangency; external/external, external/internal, and so on (Figure 3). This is well known in the 2-d case; finding solution circles for all possible tangency relationships is the celebrated 10$^{th}$ problem of Apollonius [GR04, WeisAP, KKS02]. The methods of this paper can be applied to that problem and its d-dimensional generalization in a particularly straightforward way [FN3]. Other problem dependent use and selection of vertices is discussed further in Sections III and IV.

After the evaluation of inner products, (10) and (11) contain only scalar quantities and are therefore independent of d, ensuring that the evaluation and analysis of roots is the same in all dimensions. Interestingly then, the "normal" solutions of Equation (1), those for which det($\underline{\mathbf{V}}$) ≠ 0 and the discriminate of (11) is not negative, do not depend on the dimensionality of the problem, neither in number nor in complexity. In this regard the Apollonius vertices are analogous to the Voronoi and power diagrams. On the other hand, the situations leading to no solutions are different, more numerous, and more complex than for the other diagrams (Section III). Visualization of solutions in higher than 3-d is, of course, a problem, but for 3-d, compare Figure 4a-c to the corresponding 2-d cases in Figure 3a-c (also see [MVLG06]).

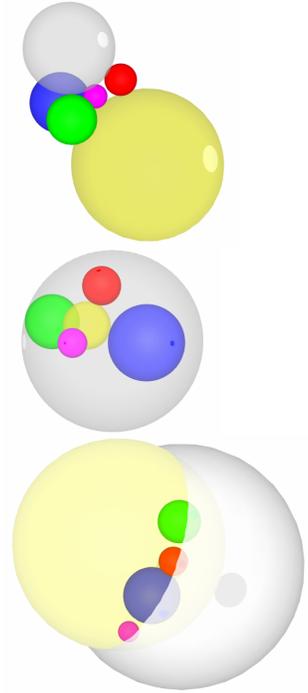

*Figure 4: Solutions-3d (yellow/white); (a-top), (b-middle),(c-bottom)*

### D. Observations on the Lifted Vector Solution

Since both lifted solution vectors $\boldsymbol{\alpha}_+$ and $\boldsymbol{\alpha}_-$ are related to the same lifted unit vector, $\hat{\boldsymbol{\alpha}}$, by scaling, it is clear the straight line in $\mathbf{R}^{d+1}$ defined by their termini also includes the origin, which above was chosen as the power vertex, p, of $\mathbf{B}$ in (3). It follows that the same is true of the projected solution vectors, $\mathbf{v}_-$ and $\mathbf{v}_+$, in $\mathbf{R}^d$. Accordingly, we define the line, $\mathbf{L_a}$, in $\mathbf{R}^d$, that runs through the termini of the two solution vectors $\mathbf{v}_-$ and $\mathbf{v}_+$, and refer to $\mathbf{L_a}$ as the Apollonius line. We can now restate the above remark without reference to an origin choice:

Observation 1: The power vertex of the ball set $\mathbf{B}$ lies on the Apollonius line, $\mathbf{L_a}$, of $\mathbf{B}$.

As in Section I.A, consider a second set of balls in $\mathbf{R}^d$, $\mathbf{B'}$, related to the balls of $\mathbf{B}$ such that for every ball $b_i$ = ($v_i$, $r_i$) ∈ $\mathbf{B}$, there is a ball $b'_i$ = ($v_i$, $r_i$ + $r_\varepsilon$) ∈ $\mathbf{B'}$, $r_\varepsilon$ ∈ $\mathbf{R}$. Here, the centers of the balls of $\mathbf{B'}$ are identical to those in $\mathbf{B}$, but each has a radius incremented by an amount, $r_\varepsilon$, relative to the corresponding ball in $\mathbf{B}$. As discussed above, such an incrementation does not produce new solution vectors. Therefore the two ball sets, $\mathbf{B}$ and $\mathbf{B'}$ share the same Apollonius line, $\mathbf{L_a}$. However, the power diagram is not invariant to radius incrementation (see Section II), so the power vertex of ball set $\mathbf{B'}$, say $\mathbf{p'}$, is not in general equal to $\mathbf{p}$ for set $\mathbf{B}$, but by the same reasoning that led to Observation 1, $\mathbf{p'}$ must also fall on the line $\mathbf{L_a}$. The above can be summarized in the form of observation:

Observation 2: The power vertices, p and p', of the ball sets defined above, $\mathbf{B}$ and $\mathbf{B'}$, are sufficient to determine the Apollonius line, $\mathbf{L_a}$, of $\mathbf{B}$ (or $\mathbf{B'}$).

### E. Scaled Lifted Vector Solution

In view of Observation 2, the line $\mathbf{L_a}$ can be considered determined once the position of a second power vertex, p', of the incremented ball set, $\mathbf{B'}$, has been calculated. Let $\hat{\mathbf{a}}$ be the unit vector in $\mathbf{R}^d$ in the direction of the line, $\mathbf{L_a}$, $\hat{\mathbf{a}}$ = ($\mathbf{p'}$ - $\mathbf{p}$)/|$\mathbf{p'}$ - $\mathbf{p}$|, and rewrite (4b) as follows:

$$\boldsymbol{\alpha} = \lambda \hat{\mathbf{a}} + r\mathbf{e}_{(d+1)} \qquad (13)$$

where we have replaced, $\mathbf{v}$, in (4b) with $\lambda \hat{\mathbf{a}}$, $\lambda \in \mathbf{R}$, to introduce the direction of $\mathbf{L}_a$. From (13) we see that the task of determining $d + 1$ unknown quantities is reduced to the determination of just two scalars, $\lambda$ and $r$. Substituting (13) into (7) furnishes a relationship between the two:

$$\hat{\mathbf{a}} \cdot \boldsymbol{\alpha}_{ik} = \lambda \mathbf{v}_{ik} \cdot \hat{\mathbf{a}} + r_{ik} r = 0 \qquad (14)$$

The ratio, $h = \lambda/r$, can be determined from (14) using any of the $d$ independent difference vectors $\boldsymbol{\alpha}_{ik}$, so long as the radius difference, $r_{ik} \neq 0$ and $\hat{\mathbf{a}} \neq 0$. If all balls $b_i \in \mathbf{B}$ have zero radius or if they all have the same radius, the Apollonius vertex is identical to the Voronoi vertex (Section II). Such cases are common in materials applications, but below it is temporarily convenient to require that there is at least one ball pair, say $b_i$ and $b_j$, such that $r_i \neq r_j$; this restriction is removed in Section II. We can then write,

$$h = \lambda/r = -r_{ik}/(\mathbf{v}_{ik} \cdot \hat{\mathbf{a}}) \qquad (15)$$

With $h$ determined as above, we can define a scaled lifted vector, $\tilde{\boldsymbol{\alpha}}$, related to $\boldsymbol{\alpha}$ but which depends only on $h$:

$$\tilde{\boldsymbol{\alpha}} = \boldsymbol{\alpha}/r = h\hat{\mathbf{a}} + \mathbf{e}_{d+1} \qquad (16)$$

Substitution of $\boldsymbol{\alpha} = r\tilde{\boldsymbol{\alpha}}$ into (5) and the constraint, $|\mathbf{v}_i|^2 - r_i^2 = r_p^2$, result in,

$$(h^2 - 1)r^2 - 2\boldsymbol{\alpha}_i \cdot \tilde{\boldsymbol{\alpha}} r + r_p^2 = 0 \qquad (17)$$

where the identity $|\mathbf{v}|^2 - r^2 = r^2(h^2 - 1)$ has been used. The solutions to (17) are:

$$r_\pm = \{\boldsymbol{\alpha}_i \cdot \tilde{\boldsymbol{\alpha}} \pm [(\boldsymbol{\alpha}_i \cdot \tilde{\boldsymbol{\alpha}})^2 - (h^2 - 1)r_p^2]^{(1/2)}\}/(h^2 - 1) \qquad (18)$$

$$\mathbf{v}_\pm = r_\pm h \hat{\mathbf{a}} \qquad (19)$$

and the position vectors of the solution balls in the original coordinate system:

$$\mathbf{x}_\pm = \mathbf{p} + \mathbf{v}_\pm \qquad (20)$$

The root analysis is identical to that given under (12), in particular, since $r_+$ and $r_-$ can both be positive or negative, the notation, $r_+$ and $r_-$, only indicates which root of (18) is being singled out. The analysis of special cases is deferred to Section II.C. Moving from (1) to (7) required calculation of the power vertex of $\mathbf{B}$, and proceeding to (13) required calculation of a second, that of $\mathbf{B'}$. In the next section the connection between power vertices in $\mathbf{R}^d$ and corresponding Apollonius vertices in $\mathbf{R}^d$ is examined more closely.

## II. Power Vertices of Ball Sets Related by Radius Incrementation

### A. Power Vertex of Ball Set **B**

The power distance, $t_{ip}^2$, between a ball $b_i = (x_i, r_i)$ in $\mathbf{R}^d$ and a point, $p$, is defined as follows:

$$t_{ip}^2 = |\mathbf{x_i} - \mathbf{p}|^2 - r_i^2 \qquad (P1)$$

If the point, p, lies outside the ball, $b_i$, then $t_{ip}$ has a simple geometric interpretation, it is the length of the tangent line from p to $b_i$. The power distance, $t_{ip}^2$, is negative if p lies inside of $b_i$ and although no real tangent line exists the power distance remains well defined. If the power distances from the point p to each of the balls in **B** are equal, the left hand side of all d + 1 equations like (P1) are identical and can be expanded and rearranged as follows:

$$\mathbf{x_i} \cdot \mathbf{p} = (|\mathbf{x_i}|^2 - r_i^2)/2 + (|\mathbf{p}|^2 - r_p^2)/2 \qquad (P2)$$

where, the constant, $r_p^2$, replaces $t_{ip}^2$, to enforce the constraint of equal power distances $t_{ip}^2 = t_{kp}^2$, $1 \le i, k \le d + 1$. If there exists a point, p, and constant $r_p^2$, that satisfy the d + 1 equations like (P2), the point p is the power vertex of **B** and at power distance, $r_p^2$. Each vertex of Figure 2b is such a point p for some set **B** and the radius of the associated power ball is $r_p$. The equation resulting from subtracting any two like (P2) eliminates the second quantity in parenthesis with the result:

$$\mathbf{x_{ik}} \cdot \mathbf{p} = (|\mathbf{x_i}|^2 - r_i^2)/2 - (|\mathbf{x_k}|^2 - r_k^2)/2 \qquad (P3)$$

where $\mathbf{x_{ik}} = \mathbf{x_i} - \mathbf{x_k}$. There are at most d such linearly independent equations like (P3) which can be used to solve for the d components of **p**. Any position vector, $\mathbf{x_i}$, $1 \le i \le d + 1$, could be taken as the kth, but the particular choice, k = d + 1, simplifies the indexing of matrices below. With, k = d + 1, define the matrix element $V_{ij}$ of the d x d matrix, $\underline{\mathbf{V}}$, as the jth component of the ith difference vector, $\mathbf{x_{i,d+1}}$, with $1 \le i, j \le d$ (*i.e.* $V_{ij} = x_{ij} - x_{(d+1)j}$). Define the ith element of the d x 1 column matrix, $\underline{\mathbf{t}}$, as the scalar quantity, $(|\mathbf{x_i}|^2 - r_i^2)/2 - (|\mathbf{x_{d+1}}|^2 - r_{d+1}^2)/2$, and let the ith element of the d x 1 column matrix, $\underline{\mathbf{p}}$, be the ith component of the vector, **p**, $1 \le i \le d$. The entire system of d equations is then represented by the matrix equation

$$\underline{\mathbf{V}}\underline{\mathbf{p}} = \underline{\mathbf{t}} \qquad (P4)$$

with the formal solution:

$$\underline{\mathbf{p}} = \underline{\mathbf{V}}^{-1}\underline{\mathbf{t}} \qquad (P5)$$

If $\det(\mathbf{V}) \ne 0$, **p**, is uniquely determined by (P5) and the power distance, $r_p^2$, from p to each ball in **B** can be calculated by rearranging (P2), where index, i, can be taken for any ball $b_i \in \mathbf{B}$. The power vertex lies outside all of the balls in **B** if $r_p^2 > 0$, if $r_p^2 < 0$, it lies within all of the balls of **B**, and $r_p^2 = 0$ represents the case where p lies on the surface of all the balls simultaneously.

In Section I we take p as the origin of coordinates and define $\mathbf{v_i} = \mathbf{x_i} - \mathbf{p}$. In the form, $\mathbf{x_i} = \mathbf{v_i} + \mathbf{p}$, we substitute this into (P2) and upon expansion and rearrangement we obtain:

$$|\mathbf{v_i}|^2 - r_i^2 = r_p^2,$$

which is just the constraint used to proceed from (6) to (7).

Here and for the remainder of this report, we use the inverse matrix in (P5) as a metaphor for any method of solving a set of linear equations. In practical numerical work the inverse matrix is rarely constructed or used since faster and more accurate methods are available and routine; for the development of predicates, Cramer's rule is often applied to obtain an explicit expression for each component of **p**, as a ratio of determinants (Section III).

Before proceeding, it is important to make a connection between the Voronoi vertex and the power vertex. Express the column matrix, $\underline{t}$, the elements of which are, $(|\mathbf{x_i}|^2 - r_i^2)/2 - (|\mathbf{x_k}|^2 - r_k^2)/2$, as the sum of two column matrices, $\underline{t}'$ and $\underline{t}''$, such that the elements of $\underline{t}'$ are $(|\mathbf{x_i}|^2 - |\mathbf{x_k}|^2)/2$, and the elements of $\underline{t}''$ are $(r_k^2 - r_i^2)/2$, so with k = d + 1 as above, (P5) takes the form:

$$\underline{p} = \underline{V}^{-1}\underline{t}' + \underline{V}^{-1}\underline{t}''$$

This shows that even if the matrix $\underline{t}''$ is null (*i.e.* all radii of **B** are equal or zero) the power vertex remains well defined. It follows that the use of power vertex as the origin of coordinates, as done in (3), does not admit special cases insofar as the radii of **B** are concerned. Likewise, once **p** is determined, the value $r_p^2$ is well defined and can be calculated from (P2) using any of the d + 1 balls in **B**, even those, if any, that have zero radii. It is also easy to show the direct calculation of the Voronoi vertex results if one takes as null the column matrix $\underline{t}''$. This justifies the statement, made below (2) that if all the radii of ball set **B** are zero, or all radii are equal, the power vertex, **p**, is identical to the Voronoi vertex.

If the difference vectors, $\mathbf{x_{ik}}$, are not linearly independent, that is, do not span d-dimensional space, det(**V**) = 0, and there is no solution of (P4) for the power vertex of **B**. Examples include three co-linear points in 2-d or four co-planar points in 3-d. For power and Voronoi diagrams this simply means the vertex doesn't exist or equivalently, that the Voronoi/power ball has a formally infinite radius. The consequences for the Apollonius diagram can be somewhat different as discussed in Section III.

## B. Radius Incrementation

As in Section I, let **B'** be the set of balls related to the set **B** such that for every ball $b_i = (x_i, r_i)$ in **B** there is a ball, $b'_i = (x_i, r'_i)$ in **B'** with $r'_i = r_i + r_\varepsilon$, with $r_\varepsilon$ the incrementation radius. By the same process as above, we obtain the following version of (P2) for the power vertex **p'** and power distance $r_p^2$ of the ball set **B'** by simply adding a prime to all quantities except $\mathbf{x_i}$ because the positions of the balls in **B'** are the same in those of **B**:

$$2\mathbf{x_i} \cdot \mathbf{p'} = (|\mathbf{x_i}|^2 - r_i'^2) + (|\mathbf{p'}|^2 - r_p'^2) \tag{P6}$$

or in terms of the radii of the original set, **B**,

$$\mathbf{x_i} \cdot \mathbf{p'} = (|\mathbf{x_i}|^2 - r_i^2)/2 - r_\varepsilon r_i - r_\varepsilon^2/2 + (|\mathbf{p'}|^2 - r_p'^2)/2 \tag{P7}$$

The difference equations corresponding to (P3) take the form

$$\mathbf{x_{ik}} \cdot \mathbf{p'} = (|\mathbf{x_i}|^2 - r_i^2)/2 - (|\mathbf{x_k}|^2 - r_k^2)/2 - r_\varepsilon r_{ik} \tag{P8}$$

where $r_{ik} = r_i - r_k$. Defining the ith element of the d x 1 column matrix $\underline{r}$, as the scalar quantity $r_i - r_{(d+1)}$, and proceeding as above we arrive at the matrix equation:

$$\underline{V}\underline{p}' = \underline{t} - r_\varepsilon \underline{r} \tag{P9}$$

where the matrices $\underline{V}$ and $\underline{t}$ are identical to those in (P4) and, $\underline{p}'$ is a column matrix with elements equal to the components of the power vertex **p'**. The power vertex of the set **B'** can now be written

$$\mathbf{p'} = \underline{\mathbf{V}}^{-1}\underline{\mathbf{t}} - r_\varepsilon \underline{\mathbf{V}}^{-1}\underline{\mathbf{r}} = \mathbf{p} - r_\varepsilon \underline{\mathbf{V}}^{-1}\underline{\mathbf{r}} \tag{P10}$$

where the substitution of **p** follows from (P5). Define the dimensionless vector, $\tilde{\mathbf{p}}$, in $\mathbf{R}^d$, to have components equal to the elements of the column matrix, $-\underline{\mathbf{V}}^{(-1)}\underline{\mathbf{r}}$:

$$\tilde{\mathbf{p}} = -\underline{\mathbf{V}}^{-1}\underline{\mathbf{r}} \tag{P11a}$$

and (P10) can be written as a vector equation,

$$\mathbf{p'} = \mathbf{p} + r_\varepsilon \tilde{\mathbf{p}}, \tag{P11b}$$

recognizable as the parametric equation of a straight line in $\mathbf{R}^d$ in the direction of the vector $\tilde{\mathbf{p}}$, which we will call the power gradient of the ball set **B**. We define the power line, $\mathbf{L_p}$, of **B** as the line in the direction of $\tilde{\mathbf{p}}$, that contains the points p and p'. Every ball set defined by uniformly incrementing the radii of the reference set, **B**, by a new value of $r_\varepsilon$, will fall on the line, $\mathbf{L_p}$. Recall from Section I that both p and p' fall on the Apollonius line, $\mathbf{L_a}$. The next observation then follows:

Observation 3. The Apollonius line, $\mathbf{L_a}$, of **B** and the power line, $\mathbf{L_p}$, of **B** are co-linear.

That the power line, if it exists, must have this property can also be deduced from Observation 2, for example, by considering the position of the power vertex of a third ball set, **B"**, incrementally related to both **B** and **B'**.

## C. Apollonius Vertices in Terms of the Power Gradient

We are now in a position to make another connection between the power diagram and the Apollonius diagram. Express the power vertex of **B'** relative to the power vertex of **B**, by defining **p"** = **p'** − **p**, and substitute into (P8) for **p'** with the result,

$$\mathbf{x_{ik}} \cdot \mathbf{p"} + \mathbf{x_{ik}} \cdot \mathbf{p} = (|\mathbf{x_i}|^2 - r_i^2)/2 - (|\mathbf{x_k}|^2 - r_k^2)/2 - r_\varepsilon r_{ik} \tag{P12}$$

Reference to (P3) shows that the quantity closest to the equal sign on the left of (P12) is equal to the first two on the right, leaving:

$$\mathbf{x_{ik}} \cdot \mathbf{p"} = -r_\varepsilon r_{ik} \tag{P13}$$

and similarly substituting **p"** for, **p'** - **p**, after rearranging (P11b),

$$\mathbf{p"} = r_\varepsilon \tilde{\mathbf{p}} \tag{P14}$$

Now consider substitution of (P14) into (P13) for **p"**:

$$\mathbf{x_{ik}} \cdot \tilde{\mathbf{p}}/r_{ik} = -1 \tag{P15}$$

comparing to (P15) to (15) and noting that $\mathbf{v_{ik}} = \mathbf{x_{ik}}$ we find:

$$\mathbf{a} = \tilde{\mathbf{p}}/|\tilde{\mathbf{p}}|, \quad h = |\tilde{\mathbf{p}}| \tag{P16}$$

This allows us to lift the restriction set down in Section I.E that $\mathbf{B}$ must contain at least one pair of balls, i and j, such that $r_i \neq r_j$. From (P11a), all such cases result in, $\tilde{\mathbf{p}} = 0$, the solution vertex is identical to the power vertex, and (P19d) below can be used to calculate the solution radius. Equation (17) can be rewritten entirely in terms of the power gradient, the power radius and a single ball from $\mathbf{B}$:

$$(|\tilde{\mathbf{p}}|^2 - 1)r^2 - 2(r_i + \mathbf{v_i} \cdot \tilde{\mathbf{p}})r + r_p^2 = 0 \tag{P17}$$

The solution radii and solution vectors are then

$$\mathbf{v}_\pm = r_\pm \tilde{\mathbf{p}} \tag{P18}$$

$$r_\pm = \{(r_i + \mathbf{v_i} \cdot \tilde{\mathbf{p}}) \pm [(r_i + \mathbf{v_i} \cdot \tilde{\mathbf{p}})^2 - r_p^2(|\tilde{\mathbf{p}}|^2 - 1)]^{(1/2)}\}/(|\tilde{\mathbf{p}}|^2 - 1) \tag{P19a}$$

Aside from the case of a negative discriminate, discussed in Section III, these equations exhibit a number of special cases. Any proper numerical treatment of the quadratic will handle all of them automatically but they are given here for completeness. The first is, $r_p^2 = 0$, the case of all generator balls intersecting at a point, the second, $|\tilde{\mathbf{p}}| = 1$, is the case of all balls sharing supporting tangent line (2d), plane (3d), or hyperplane (d > 3), the third, $|\tilde{\mathbf{p}}| = 0$, is the case of all balls with equal radii. The resulting simplifications of (P19a) are as follows, where any ball can be taken as the ith:

$$r_p^2 = 0, \; |\tilde{\mathbf{p}}| \neq 1: \qquad r_+ = 0, \; r_- = 2(r_i + \mathbf{v_i} \cdot \tilde{\mathbf{p}})/(|\tilde{\mathbf{p}}|^2 - 1) \tag{P19b}$$

$$|\tilde{\mathbf{p}}| = 1, \; (r_i + \mathbf{v_i} \cdot \tilde{\mathbf{p}}) \neq 0: \qquad r = (1/2)r_p^2/(r_i + \mathbf{v_i} \cdot \tilde{\mathbf{p}}) \tag{P19c}$$

$$|\tilde{\mathbf{p}}| = 0: \qquad r = -r_i \pm (r_i^2 + r_p^2)^{(1/2)} \tag{P19d}$$

If $\det(\underline{\mathbf{V}}) = 0$, then as with power vertex, $\mathbf{p}$, the power gradient, $\tilde{\mathbf{p}}$, cannot be calculated (see P11a) and (P18-22) are no longer valid. This condition is then detected in (P4, P5) where the power vertex is calculated; discussion of the $\det(\mathbf{V}) = 0$ case is deferred to Section III.

In terms of the original coordinate system, the solutions are as follows:

$$\mathbf{x}_\pm = \mathbf{p} + \mathbf{v}_\pm = \mathbf{p} + r_\pm \tilde{\mathbf{p}} \tag{P20}$$

$$r_\pm = \{(r_i + (\mathbf{x_i} - \mathbf{p}) \cdot \tilde{\mathbf{p}}) \pm [(r_i + ((\mathbf{x_i} - \mathbf{p}) \cdot \tilde{\mathbf{p}})^2 - r_p^2(|\tilde{\mathbf{p}}|^2 - 1)]^{(1/2)}\}/(|\tilde{\mathbf{p}}|^2 - 1) \tag{P21}$$

Suppose, the generator balls of $\mathbf{B}$ have been preprocessed such that $x_i = x'_i - x'_k$, and $r_i = r'_i - x'_k$, where the primed quantities are actually the original set. Then for the kth ball, which can be any member of the set, $x_k = 0$, $r_k = 0$, and substitution into (P21) yields:

$$r_\pm = \{-\mathbf{p} \cdot \tilde{\mathbf{p}} \pm [(\mathbf{p} \cdot \tilde{\mathbf{p}})^2 - |\mathbf{p}|^2(|\tilde{\mathbf{p}}|^2 - 1)]^{(1/2)}\}/(|\tilde{\mathbf{p}}|^2 - 1) \tag{P22}$$

where in (P22), the equality $r_p^2 = |\mathbf{p}|^2$ holds since the ball with $r_k = 0$ was taken as the origin of coordinates for the preprocessed set. With (P20) and P(22) we have complete solutions for the set $\mathbf{B}$ which involve only vector sums and inner products for two vectors, both of which are determined by the matrix $\underline{\mathbf{V}}^{-1}$ (P10). The "preprocessing" described above can be built into the algorithm if the use of (P22) is desired and, of course, implies a post-processing step to recover the values of the original variables.

Note that if the balls are preprocessed as described above, cases with small negative solutions balls (Figures 3d - 3f) are no longer possible. To understand this, consider the example in Figure 5a with two negative solutions, one large and one small (the same configuration type as in Figure 3e). In Figure 5b all generators have been decremented (*i.e.* incremented by a negative $r_\varepsilon$) by an amount equal to half the radius of the smallest generator. There is no longer a common intersection of the three generators making a small negative solution impossible and a positive solution has emerged to replace it. In Figure 5c, the decrementation process is taken to its logical conclusion and the smallest ball reduced to a point. This result is two positive solutions just as in Figure 5b. It is easy to show that if no ball of **B** is trivial, a uniform incrementation that leaves the smallest ball a point will always result in a set with no common intersection and therefore no small negative solutions. From this, one might surmise that requiring a disjoint ball set is a restriction without consequence, but Figure 5c shows otherwise since the two larger balls are still intersecting.

Apparently, the requirement of strictly disjoint ball sets limits the configurations under consideration. This can be a confusing point in the literature because on the one hand many authors illustrate only disjoint ball configurations even though the algorithms under consideration are more general, while on the other hand, there are algorithms that do require the generator set to be strictly disjoint.

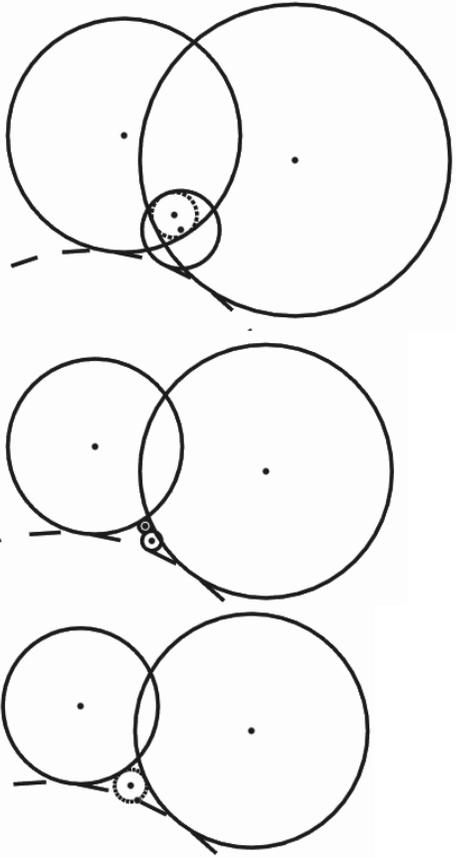

*Figure 5: Small Negative Solution Behavior;  (a-top), (b-middle), (c-bottom)*

## Section III.  Implementation

### A. Recipes

In this section we discuss various aspects of the vertex equations relevant to implementation. In terms of computational recipes for the Apollonius diagram, the vertex solutions can be summarized as follows:

Recipe 1 – Single Linear Equation Solver Call or Matrix Inversion, Eqs. (P4, P11a, P20, P21)
    a) Assemble the d x d matrix, **V**, and the d x 1 matrices **t** and **r**
    b) Calculate **p** and **p̃** from the matrix equations **Vp** = **t** and **Vp̃** = -**r**
    c) Calculate the solution radii $r_+$ and $r_-$ according to (P21)
    d) Discard imaginary roots and "large" negative radii; keep all others
    e) For each solution radius retained, calculate the solution vector $\mathbf{x}_\pm$ = **p** + $r_\pm$**p̃**

Recipe 2 – Double Power Vertex via Incrementation, Section I.E
    a) Assemble the d x d matrix **V**, and d x 1 matrix **t**
    b) Assemble the matrix **t'** based on incremented ball radii, $r_i' = r_i + 1$
    c) Calculate p and p' from **Vp** = **t** and **p'V** = **t'**
    d) If, **p'** = **p**, then,  **â** = **0**,  **x** = **p**, use (P19d) to calculate radii, see discussion under (P16)
    e) Otherwise, calculate **â** = (**p'** – **p**)/|**p'** – **p**|, h = - $r_{ik}$/($\mathbf{v_{ik}} \cdot \mathbf{\hat{a}}$) (any i, k such that $r_{ik} \neq 0$)

f) Form, $\tilde{\boldsymbol{\alpha}} = h\hat{\mathbf{a}} + \mathbf{e}_{d+1}$, calculate $r_\pm$ with (18) using any $\boldsymbol{\alpha}_i$; discard any large negative radii

g) For each radius retained, calculate a solution vector, $\mathbf{x}_\pm = \mathbf{p} + r_\pm h\hat{\mathbf{a}}$

Recipe 3 – Solutions via Normal Vector in $\mathbf{R}^{d+1}$, Eqs. (8 -12)

a) Calculate a normal vector, $\boldsymbol{\alpha}'$, as detailed in Appendix I.
b) Normalize to obtain $\hat{\boldsymbol{\alpha}}$, obeying the sign convention discussed under (11)
c) Calculate $\sigma_\pm$ to obtain $\boldsymbol{\alpha}_\pm$ according to (11) and (12)
d) Recover $\mathbf{v}_\pm$ as the $1^{st}$ d components, and $r_\pm$ as component d + 1 of $\boldsymbol{\alpha}_\pm = \mathbf{v}_\pm + r_\pm \mathbf{e}_{d+1}$
e) For each solution radius retained use the corresponding, $\mathbf{v}_\pm$, to calculate $\mathbf{x}_\pm = \mathbf{p} + \mathbf{v}_\pm$

Recipe 4 – Sub-dimensional Cases, det($\underline{\mathbf{V}}$) = 0 (see Section F and Appendix II).

In terms of efficiency and numerical stability, the use of Recipe 3, for example, in 3-d, implies the manipulation of a 4 x 4 matrix, and determinate evaluations of 3 x 3 matrices to find a normal vector in 4-d. The method is $O(d^4)$ and also uses 4 x 1 column matrices throughout. It has the highest operation count of the three, and because vector $\boldsymbol{\alpha}'$ of (8) needs to be normalized, it involves higher degree polynomials based on input. The remaining two equation sets require 3 x 3 and 3 x 1 matrices and are both $O(d^3)$. The use of Recipe 2 implies finding power vertices for both $\mathbf{B}$ and $\mathbf{B}'$ and the use of some 4-component vectors. The method envisions the availability of a power vertex routine but two calls to a linear equation solver will work as well. The value of $r_p^2$ is required and can be calculated with (P2) if it isn't returned by the power vertex routine. Also note that if all of the balls of $\mathbf{B}$ have the same radius, $\mathbf{p} = \mathbf{p}'$, $\mathbf{v} = \mathbf{p}$, and the solution radius should be calculated with (P19d). If code is already available for the calculation of power vertices, implementation requires only vector algebra to find the Apollonius vertices.

Implementation of Recipe 1 can be achieved with a single matrix inversion, or better, with a single call to a linear equation solver (*cf.* [PFTV88], Chapter 2). Since from (P4), $\underline{\mathbf{V}}\mathbf{p} = \underline{\mathbf{t}}$ and from (P10 – 11), $\underline{\mathbf{V}}\tilde{\mathbf{p}} = -\underline{\mathbf{r}}$, one can pass matrices $\underline{\mathbf{V}}$, $\underline{\mathbf{t}}$, and $-\underline{\mathbf{r}}$, into the solver and the vectors $\mathbf{p}$ and $\tilde{\mathbf{p}}$ are returned, normally with an error tolerance close to machine precision.

Recipes 1 - 3 above involve the solution of a quadratic equation which has been given in standard form in each case. In isolation, this form, or any other form of the quadratic equation, is susceptible to round-off error; a numerically stable way to solve a quadratic is given in ([PFTV88], Chapter 5) and is easily implemented.

## B. Imaginary Solutions

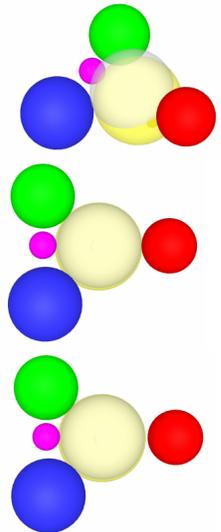

Figure 6: Imaginary Solution; (a-top),(b-middle), (c-bottom)

If the discriminate is negative, there are no real roots and therefore no solution balls, but there are a pair of complex conjugate roots which here we call imaginary roots/solutions. In any dimension this occurs when any ball of $\mathbf{B}$ is completely contained in another ball of $\mathbf{B}$; the contained ball is often called trivial. However there are ball configurations in all dimensions that give rise to imaginary roots, but have no trivial balls. An example for 2-d is given in [KY02], one for 3-d in [MVLG06], and another 3-d example by [Will99] that involves seven generators. In the latter work it is pointed out that in 3-d, such configurations involve a ball that is enclosed in the convex hull of two others. Generator balls with no vertices are said to be "disconnected" when the diagram is considered a graph and this can present various problems in the design of algorithms for diagram generation and usage. The distinction between these and trivial balls is that non-trivial disconnected balls can have a non-empty cell (*i.e.* a cell with non-zero area (2-d), volume (3-d), etc.) even though it lacks vertices entirely. Apparently, the number of such configurations possible in various dimensions is not known. However, any such configuration is detected by the presence of imaginary roots; once detected, distinguishing trivial balls from other forms of disconnection is obviously not difficult. A 3-d example of a configuration that is close to yielding imaginary solutions but actually results in two ordinary positive solutions is shown in Figure 6a. In Figure 6b, the same configuration is viewed roughly down the Apollonius line; considering both views it is clear that the smallest ball is inside the convex hull of the largest two. In Figure 6c, the red ball is moved by 1/10 of its radius away from the remaining three, the solution balls of Figure 6a/b are included for scale because the configuration of Figure 6c gives rise only to a pair of imaginary solutions. This is simply a case where the pairwise hyperbolic half spaces fail to meet at a single point; obviously such a configuration can be very close to yielding ordinary real solutions.

## C. Solution Selection

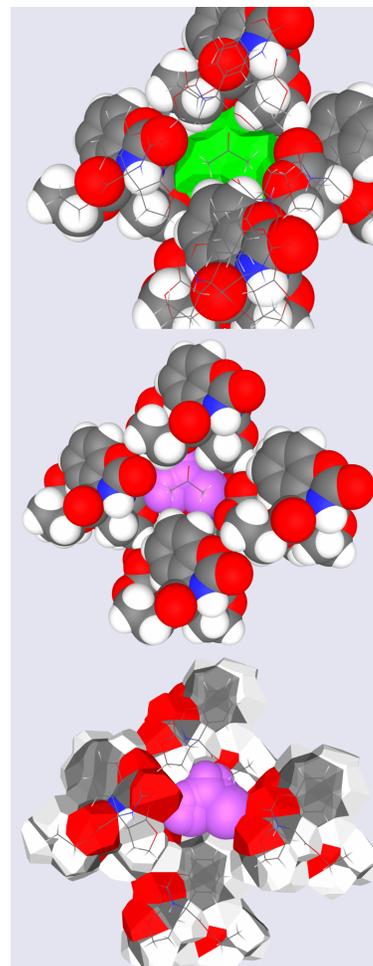

*Figure 7: Solvent Pocket (a-top), (b-middle), (c-bottom)*

As discussed in Section I.C and the recipes above, real roots with "large" negative radii (Figure 3B, c, and e) are not vertices of the Apollonius diagram. Here, by "large negative" we mean large enough to contain all the balls of **B**, but since this condition is not conveniently computed, the "large negative" condition is decided as follows: let R = |r| where r is any negative solution ball of **B**. Since the radii of all generator balls are non-negative, let Rmax be the largest radius of set **B**. Then, if R > Rmax, a negative solution ball is a "large negative" solution ball; if desired, "small negative" can be defined similarly using the smallest ball of **B**. Moreover, if one is willing to preprocess and post-process each ball set individually and use (P22) for solutions, the criterion becomes completely trivial: only positive balls can be part of the diagram. For other applications, solution selection is dictated by the problem under consideration, so Apollonius balls with "small" negative radii and even those with small positive radii may eventually or immediately be discarded. Possible examples include motion planning, medial axis calculation, and the study of voids and channels at atomic scale. Two potential examples for the use of large negative balls are mentioned in Section IV.

As an example of vertex selection and the use of the Apollonius vertices in the characterization of voids, consider the not unusual situation that a compound that crystallizes with solvent, partially or entirely loses its solvent(s) of crystallization over time. This can result in a crystal that retains the original space group, exhibits minimal changes in cell constants, but displays a crystal structure with large void space(s). We simulate this by starting with the experimental crystal structure [SNA03] of a well-behaved acetone solvate and computationally desolvating it. In Figure 7a, we show the crystal structure in the vicinity of the binding pocket with some cage molecules removed to make visible the solvent molecule [FN2]. The solvent is represented by its power cells and stick diagram, while the cage molecules are rendered as a space filling diagram (union of van der Waals spheres) with the stick diagram overlaid; note the obvious complementarity between the power faces of the solvent and the cage spheres at the surface of the pocket. After removing the solvent, Apollonius balls were calculated and a subset of these used to fill the void (Figure 7b). These vertices were arrived at as follows: first all vertices of the desolvated structure were calculated as described in Recipe 1 discarding all large negative balls;

then, the resulting vertex set was culled on radius such that only vertices with atom sized or larger radii were retained. The Apollonius balls fill more of the pocket than the original solvent but there are many more of them. In Figure 7c, the atoms of the cage molecules are rendered as their power cells and the complementarity to the Apollonius balls filling the pocket is obvious. If the compound were crystallized from an acetone/acetic acid mixture, there would be no doubt as to which solvent occupied the pocket before desolvation occurred; similarly, if from an acetone/water mixture, both solvents in the pocket is also obviously not possible. Other kinds of problems in solvation and complexation can be approached with the aid of these diagrams.

## D. Twin Solutions

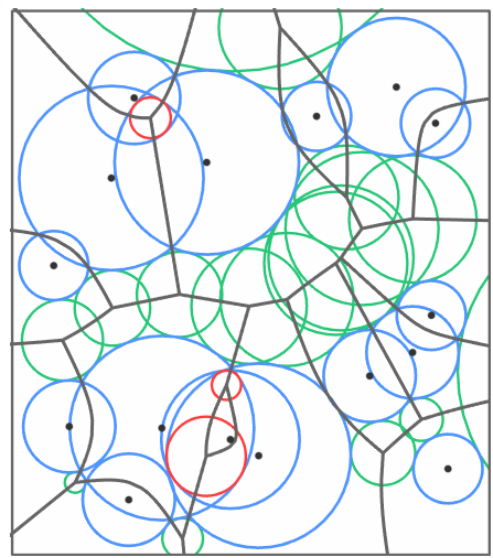

Figure 8: Apollonius Overlay with Transparent Generator Balls (same diagram as 2c): negative solutions in red, positive in green

The curved edges of the Apollonius diagram allow the formation of cells with fewer than $d + 1$ vertices, which is not possible with the straight edges of the Voronoi and power diagrams. In Figure 2c there are two examples of this, perhaps more apparent in Figure 8, one in the top right of the diagram with both vertex radii positive and one in the bottom left with both negative; in both cases a smaller ball is between two larger and it is the cell of the small ball that has two vertices rather than three. Moreover, the cell of the small ball interrupts the edge shared by the larger two; in 3-d the corresponding result is a face with a "hole" in it. A remarkable aspect of these arrangements, in all dimensions, is not only that two vertices are sufficient to define an entire cell, but that the vertices of such a cell are always the two roots of the same quadratic equation. The twin solution radii can be either both negative, both positive, or opposite signs. Such a cell could be said to be "twin bound" and has 2 edges (2-d), 3 faces (3-d), and so on. Since any ball configuration that gives rise to two positive solutions results in such a cell, Figure 4a is therefore a 3-d example. Any algorithm that calculates the entire Apollonius diagram and that is also designed to be vertex aware, can have no difficulties in detecting or tracking twin bound cells as their presence is clearly evident at the time the parent quadratic is solved.

## E. Numerical Precision and Exact Methods

The equations presented in this work are exact in the usual sense that no approximation has been made in their derivation. Some applications are deemed to require extremely high numerical precision or even exact methods. These ends can be accomplished by using multi-precision numerical types that guarantee that some set of arithmetic operations return either exact values or exact comparisons in the second case, or, values of at least some prespecified precision in the first. In this context, "exact" irrational numbers obviously have a different meaning than that normally used for integers and rational numbers. The development of such numerical types is a research area in itself and we refer the reader to [Shew96, Shew96b, MS01, KLPY99] as entry points. To use such a type, one needs to write/modify each routine in the algorithm that uses arithmetic to use the numeric type of choice instead of an ordinary, "built-in" floating point type. The possibility of selecting a particular numeric type from a "menu" of such types as a run time option depends on language facilities; in C++ the templating facility will do the job. With or without templating, it is no more difficult to implement the vertex solution methods reported here using a multi-precision type than for a "built-in" type.

Typical exact applications, rather than calculating and storing a numerical representation of geometric objects, use multi-precision types to evaluate "exact predicates", often resulting in a true/false answer which encodes the sign of an evaluated expression. For use with vertex calculations, the predicate of interest is the "incircle" predicate. For this, one would use (1) in the form:

$$|\mathbf{x}_q - \mathbf{x}|^2 - (r_q + r)^2 < 0$$

where query ball, $b_q$, is not a member of **B**, and ball $b = (\mathbf{x}, r)$ is a solution vertex of **B**. If the sign of the left hand side is negative, then the distance between the center of the query ball and the center of the solution ball is less than the sum of their radii, the two are said to be "in conflict", and the value of the predicate is "true". Upon expanding the above, one first uses explicit expressions for **x** and r, (e.g. (P20) and (P21)), to assess the algebraic degree of the entire expression and finally the precision required to determine the sign of the expression. In such analyses it is customary to assume that the input quantities are integers, so here we do the same. Examples of such an analysis in the context of the Apollonius vertices can be found in [NTS07] and [NS08] while detailed examples of evaluating the signs of non-trivial expressions are given in [EK06]. Here, we simply show that the vertex solutions of this work are readily harnessed for such an analysis. Expansion of the above will contain the vectors **p** and **p̃** so an explicit expression for each is required. From (P4) we have $\underline{\mathbf{V}}\mathbf{p} = \mathbf{t}$. We let $\underline{\mathbf{V}}_i$ represent the matrix that is identical to, **V**, except that its $i^{th}$ column is replaced by **t**. Then, denoting determinants by |…|, from Cramer's rule, the ith component of **p** can be written:

$$p_i = |\underline{\mathbf{V}}_i|/|\underline{\mathbf{V}}|, \quad |\underline{\mathbf{V}}| \neq 0.$$

Matrix $\underline{\mathbf{V}}$ contains only input coordinate differences as elements while $\underline{\mathbf{V}}_i$ contains these and in addition, $2^{nd}$ degree terms from **t**. Since the input balls are all presumed to have integer coordinates and radii it follows that both $|\underline{\mathbf{V}}_i|$ and $|\underline{\mathbf{V}}|$ are integers and therefore each component of **p** is rational. A Laplace expansion of the determinants shows that $|\underline{\mathbf{V}}|$ is degree d on input quantities and that $|\underline{\mathbf{V}}_i|$ is degree d + 1. Analysis for the vector **p̃** is strictly analogous. We also need an expression for, r, and from the above and (P21), it is clear that the radical contains only integer and rational quantities, so that the sign of the left hand side of the inequality above can be evaluated exactly even when the value of r cannot. The entire process is quite laborious and a complete analysis of the predicate is not given here.

F. Solutions for Sub-dimensional Cases

Since all methods presented use the power vertex of **B** as the origin of coordinates, they all rely on the non-singularity of matrix, $\underline{\mathbf{V}}$, defined above (4). Note that det($\underline{\mathbf{V}}$) does not involve the radii of **B** so this is strictly a consequence of the configuration of the ball centers in space. In the d = 2 case (*i.e.* circles on a plane), det($\underline{\mathbf{V}}$) = 0 only occurs if the three centers of **B** lie on a straight line. For d = 3 (four spheres in space) all four sphere centers must lie in a plane which includes the possibility that all four lie on a straight line. Such configurations result in a set of d difference vectors, $x_i - x_k$, $1 \leq i \leq d$, $k = d + 1$, that are not linearly independent and hence do span $\mathbf{R}^d$. Such cases are termed "sub-dimensional" here but in the literature, most often they are simply violations of the "all sites in general positions" restriction (not to be confused with the crystallographic notion of a general position) or sometimes referred to as "degenerate", which can include other inconvenient configurations as well. Numerically, det(**V**) = 0, implies no larger than roundoff error. Geometrically, this means that extremely small deviations from linearity for (2-d) or from planarity

(3-d) result in a non-zero determinant and a numerical solution for the power vertex.

Based on limited experience with randomly generated sites and experimental data as input, cases with det($\underline{V}$) strictly zero are probably somewhat rare as they have not been detected. Nevertheless, they are clearly possible and a complete implementation needs to handle them. Perturbation methods can be applied without significant alteration of the method. For example, if the input balls are from experimental data, the coordinates can be perturbed by an amount small compared to experimental error but sufficiently large that numerically, det($\underline{V}$) ≠ 0. Presumably, it is also possible to adapt methods similar to those usually applied to predicates [EM90, EC95], to produce an arbitrarily small perturbation. Either way, the matrix would be non-singular and the power vertex and the solution vertices calculable.

Although perturbation approaches are conventional and practical they may not be entirely satisfactory in the case of the Apollonius solutions. Sub-dimensional cases are not difficult to construct, nor to solve, and without approximation. Will reports a method for cell extraction in 3-d [Will99] that handles such configurations but does not report vertex solutions. In Appendix II, we give algebraic solutions for all cases of det($\underline{V}$) = 0 for which rank($\underline{V}$) = d – 1 (Recipe 4). These should be the most common sub-dimensional cases in the sense they are realized with the fewest geometric constraints, and apparently, only these yield discrete solutions. As a 3-d example, the sub-dimensional configuration in Figure 9a has all ball centers in the xy plane and the solution balls lie symmetrically disposed above and below the plane with equal radii. All coordinates and generator radii are order of unity as are both the solution vector moduli and the solution radius. All calculations were performed with the C++ floating point type, `double;` the machine precision for this type is approximately 1 x $10^{-16}$ for CPU/OS/compiler combination used in this study. Instead of a linear equation solver, an ordinary matrix inverter was used to encourage a rapid accumulation of round-off error. The absolute error in both the solution vector moduli (hereafter |Δ$\mathbf{x}$|) and radius, |Δr|, using Recipe 4, is nonetheless still close to machine precision at 1 x $10^{-16}$. For an approximate solution using a finite perturbation approach, an obvious idea is to nudge a single ball slightly out of plane and use Recipe 1; here we choose the ball at the origin for perturbation and denote the nudge as Δz. Taking Δz = 1 x $10^{-7}$, results in |Δ$\mathbf{x}$| of 5 x $10^{-6}$ and |Δr| of 1.5 x $10^{-5}$. For many purposes this could well be sufficiently precise, but choosing Δz an order larger, incurs errors an order larger as well, and so on. Can we do better by taking Δz smaller? Figure 9b shows the result of taking Δz = 4.4 x $10^{-10}$; the resulting |Δr| is only $10^{-7}$, but the moduli are obviously far from correct with errors in |Δ$\mathbf{x}$| in excess of 40%; with even smaller Δz, solution quality continues to deteriorate. There appears to be an optimal perturbation in the sense that larger perturbations lead to higher radius errors while smaller ones result in larger vector errors. This suggests that it is well worth the extra effort to code Recipe 4 and use it whenever $\underline{V}$ is singular and perhaps even when det($\underline{V}$) is "sufficiently small" for a given application. For exact algorithms, predicates based on solutions from Recipe 1 will be different than those based on Recipe 4. Predicate selection can be based entirely on determination of det($\underline{V}$) and both predicates appear to be necessary for a complete implementation.

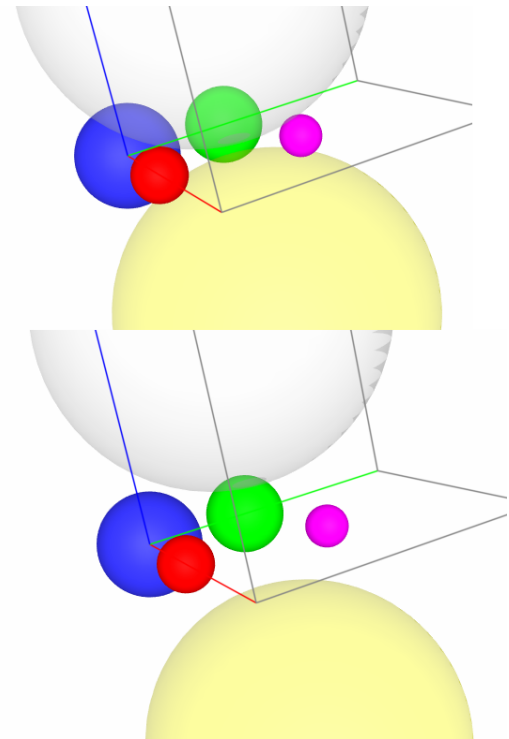

*Figure 9: Sub-dimensional and Perturbation Solutions/gray/yellow; (a-top), (b-bottom)*

Also note, that the motivation for taking the power vertex in $\mathbf{R}^d$ as the origin of coordinates was to render

(6) homogeneous; it follows that any origin choice that results in (7) could be used instead. This would lead to a modified version of the matrix, **A**, (Appendix I), the determinate of which would not necessarily be zero even if det|**V**| (*i.e.* the upper left d x d of **A**) is precisely zero. Such a matrix could then provide solutions, discrete and otherwise, even when the power vertex in $\mathbf{R}^d$ does not exist. We hope to return to this point in a future report.

## G. Related Work

We are aware of only one other report of Apollonius solutions in $\mathbf{R}^d$, that of Gavrilova [Gav09]. As with the present work, the choice of notation directly supports coding vector and matrix algorithms. The method of solution diverges from the present one almost immediately because the choice of origin is the center of one of the balls of **B**. In the general case, this leads to Cramer's rule solutions such that each component of each solution vector is a function of the yet undetermined solution radius. Solution vector calculations then require a series of back substitutions, d in number, after a radius has been calculated. Precision control is implemented by Newton's method, utilizing the full-matrix Jacobian; precision is improved iteratively and this is reported to be the most time consuming step of the algorithm. With the methods reported here, precision control amounts to calculating the value of a single scalar, which is in general irrational, with sufficient precision to support its use as a scale factor in (12) or (P20). For example, in (P20), the vectors $\mathbf{v}_i$, **p**, and **p̃** are known, either exactly or to any required precision, in advance of the radius calculation of (P21). The precision required for evaluation of the radical can therefore be assessed before it is actually calculated which does not require an iterative procedure.

## Section IV. Discussion

Viewed from the power vertex in $\mathbf{R}^d$, the lifted solution vectors are readily characterized as being normal to the lifted difference vectors of the generators. This not only leads to a solution method (Recipe 3) but the form of the solution (12) makes it clear that the power vertex must fall on the line defined by the termini of the projected solution vectors (here called the Apollonius line), as stated in Observation 1 of Section I. We find no mention of this in the literature, even for the 2-d case, where the power vertex has been used in finding Apollonius solutions at least since the work of Gergonne (*cf.* [GR04]). It is perhaps obvious that the power vertices of incremented ball sets fall on a straight line, but the fact that this line is identical to the Apollonius line (Observation 3 of Section II) was similarly not found in the literature. This leads to alternative methods of finding the Apollonius line and to new solution methods (Recipes 1 & 2).

The Apollonius diagram then has a reputation for complexity that is not entirely deserved. Vertex solutions can proceed just as for Voronoi and power vertices but the result is two vectors instead of one; both can be calculated with standard matrix tools, multi-precision arithmetic, or used in the design of exact predicates. Calculation of vertex radii is also similar to the other diagrams, although for the Apollonius diagram the solution radius is used as a scale factor for solution vectors as well. Computing the roots of the required quadratic equation is more an opportunity than a complication. It should be compared to the alternative of cataloging all possible ball configurations and devising a potentially different solution algorithm for each. It was also shown that the real roots relevant to the diagram are readily distinguished from those that are not (*i.e.* large negative balls). Secondly, trivial and disconnected balls both produce imaginary roots and distinguishing which is the case is also not difficult. Thirdly, "twin bound" regions, (*i.e.* regions with only two vertices) occur whenever two real roots of a single quadratic are both diagram relevant, rendering these configurations easily detected and tracked.

Unlike the Voronoi and power diagrams, sub-dimensional ball arrangements do not necessarily rule out the occurrence of solutions for the Apollonius diagram. To the best of our knowledge, solutions for sub-dimensional cases in $\mathbf{R}^d$, like those given in Appendix II, have not been described previously. Although the solutions again generally occur in pairs, it is only the position of the vertex that differentiates the pair; a single radius applies to both solutions so either both solutions are diagram relevant or both not.

The tangency solutions presented here and variations based on signed generator radii [FN3] represent a tool of considerable versatility in the study of ball arrangements. Although a variety of applications have been pointed out above, none of them make use of large negative solutions. Here, we cite two possibilities: the point set problem called the 1-center problem (*cf.* Chapter 3, [OBSC00]), can be generalized to balls and is often referred to as the miniball problem. The problem is to find the smallest ball that encloses a set of balls [Meg83]. For this problem, a vertex characterized by a large negative radius will normally be the solution. A related problem for points, the farthest-point Voronoi diagram (*cf.* Chapter 3, [OBSC00], Chapter 6, [AKL13]) can also be generalized to balls of any dimension and possibly in more than one way. For the 2-d problem, one such generalization has been termed the "farthest-point, farthest-site" diagram [SH09]. Here also, the solution vertices normally consist of large negative balls.

## Acknowledgements

The author thanks his wife, Rebecca, for proofreading the manuscript and for support throughout this work.

## Footnotes

[FN1] The power diagram is also known as the "Voronoi diagram in the Laguerre geometry" and the term "Dirichlet domains using radical planes" can be also encountered. The Apollonius diagram is also known as the "additively weighted Voronoi diagram (AWV)", "Voronoi diagram of circles/spheres", and the "Euclidean Voronoi diagram".

[FN2] All figures in this work were prepared using the C++ program LatticeView. Downloads will become available in 2016 from LatticeSmith.com.

[FN3] To find solutions other than those where the solution ball tangency is symmetric with respect to the generators, (1) can be generalized as follows:

$$|\mathbf{x_i} - \mathbf{x}|^2 = (r + s_i r_i)^2, s_i = \pm 1, 1 \leq i \leq d + 1.$$

Each independent set of signs, $\{s_i\}$, defines a different but related tangency problem for a given generator set (*cf.* [WeisAP]). Finding the solution balls for all such sign sets in 2-d is the classical 10[th] problem of Apollonius. The problem generalizes as expected to d-dimensions and in terms of the methods described here, for each sign set of interest, an additional column of radius differences is added to the right hand side of (P10). Each column of radius differences will generate its own power gradient so that all solution vertices and radii are calculable with a single matrix inversion or call to a linear equation solver. Note that this implies that the Apollonius line for every sign set must run through the power vertex which is invariant with respect to the signs used above. This can be seen by noting that the right hand side of (P1) would be $|\mathbf{x_i} - \mathbf{p}|^2 - (s_i r_i)^2 = |\mathbf{x_i} - \mathbf{p}|^2 - r_i^2$, thus leaving (P1) unchanged. Also note, that in modern descriptions of the Apollonius' 10$^{th}$ problem [GR04, WeisAP, KKS02], it is not always clear whether the generator balls are required to be disjoint; with the methods presented here this is not required.

[FN4] The proof of Theorem 1 closely follows those by Thomas and Spiegel (see [Thom68], pg. a15, Theorem 12, and, [Spie71], pgs. 346 and 356, Theorem 15-9). It is included here to make the main methods of this report self-contained.

# Appendix I: Normal Vectors in $\mathbf{R}^{d+1}$

## A. Finding a normal vector in $\mathbf{R}^{d+1}$

In Section I, we arrive at a set of vector equations, d in number, each vector having d+1 components:

$$\boldsymbol{\alpha}_{ik} \cdot \boldsymbol{\alpha}' = 0 \qquad (N1)$$

where $\boldsymbol{\alpha}_{ik} = \boldsymbol{\alpha}_i - \boldsymbol{\alpha}_k$; there are d+1 vectors and a normal vector, $\boldsymbol{\alpha}'$, to be determined. Let $k = d + 1$ and define the $d + 1 \times d + 1$ matrix, $\underline{\mathbf{A}}$, having elements $a_{ij}$ that are the jth component of the vector $\boldsymbol{\alpha}_{i(d+1)}$, $1 \leq i \leq (d + 1)$, or in other words, rows 1 through d hold the vectors $\boldsymbol{\alpha}_{1k}$ through $\boldsymbol{\alpha}_{dk}$, with $k = d + 1$. For present purposes it is not necessary to define the elements of the row $d + 1$ of $\underline{\mathbf{A}}$ because we are only interested in its cofactors. The elements themselves can all be taken as zero or they can be thought of as symbolically representing the unit vectors, $\hat{\mathbf{e}}_1, \hat{\mathbf{e}}_2, \ldots, \hat{\mathbf{e}}_{d+1}$, parallel to the coordinate axes of an orthonormal system in $\mathbf{R}^{d+1}$. Denote by $A_{ij}$, the cofactor of the element $a_{ij}$. Further define a new matrix, $\underline{\mathbf{B}}$, with elements identical to $\underline{\mathbf{A}}$ except that in $\underline{\mathbf{B}}$, row $d + 1$ is replaced by the cofactors of the row $d + 1$ in $\underline{\mathbf{A}}$. According to Theorem 1 below, the row $d + 1$ of $\underline{\mathbf{B}}$ is then orthogonal to each of the other rows of $\underline{\mathbf{B}}$. It immediately follows that we can define a vector, $\boldsymbol{\alpha}'$, with components equal to the elements of row $d + 1$ of $\underline{\mathbf{B}}$ which are the cofactors $A_{(d+1)j}$:

$$\boldsymbol{\alpha}' = A_{(d+1),1}\mathbf{e}_1 + A_{(d+1),2}\mathbf{e}_2 + \ldots + A_{(d+1),(d+1)}\mathbf{e}_{d+1} \qquad (N2)$$

It must be true that $\boldsymbol{\alpha}_{ik} \cdot \boldsymbol{\alpha}' = 0$, for $1 \leq i \leq d$, and $k = d + 1$, because this inner product is identical to the row orthogonality condition of Theorem 2 for row $d + 1$ with any of the other rows of $\underline{\mathbf{B}}$.

The use of cofactors here to represent vector components is operationally identical to the use of a determinate for finding the cross-product of two 3-component vectors and also closely related to area calculations via determinants (*cf.* [ORou94], pg 20)

## B. Theorems and Proofs

Theorem 1 [FN4]:

If the elements of any row of a square matrix are multiplied by corresponding cofactors of any other row, the products sum to zero.

Proof:

Let the elements of the matrix $\mathbf{A}$ be $a_{ij}$ and the cofactor of the element $a_{ij}$ be $A_{ij}$. Then the Laplace development of det($\mathbf{A}$) in terms of the ith row is:

$$\sum_k a_{ik} A_{ik} = \det(\mathbf{A}) \qquad (N3)$$

Now define a new matrix, **A'**, with elements $a'_{ij}$ and cofactors $A'_{ij}$. Let the elements of **A'** be identical to those of **A** except that the ith row of **A'** is equal to the jth row of A, $j \neq i$, so that **A'** has two rows that are equal to jth in **A**, (and therefore no longer has a row equal to ith of **A**). Since **A'** contains two identical rows, any valid expression for det(**A'**) can be set to zero. The Laplace development of det(**A'**) in terms of the ith row can then be written

$$\sum_k a'_{ik} A'_{ik} = 0 \qquad (N4)$$

But according to the definition of **A'** in terms of **A**, all the cofactors $A'_{ik} = A_{ik}$ because **A'** and **A** differ only in the ith row, and for all of the elements of the ith row, $a'_{ik} = a_{jk}$. With these substitutions (N4) can be written

$$\sum_k a_{jk} A_{ik} = 0, \quad j \neq i \qquad (N5)$$

The terms on the left hand side are the products of the elements of one row with the corresponding cofactors of a different row, so (N5) is the required result.

Theorem 2:

If any row of a square matrix is replaced by its corresponding cofactors then the new row is orthogonal to all the others.

Proof: Let the elements of the square matrix **A** be $a_{ij}$ with cofactors $A_{ij}$. Define the matrix **B** with elements $b_{ij}$ and cofactors $B_{ij}$ such that the elements of **B** are identical to those of **A** except that in **B**, the elements of the ith row are set equal to cofactors of the ith row of **A**.

Since the ith row of **B** contains cofactors as elements, the orthogonality condition to be proved can be written,

$$\sum_k b_{jk} b_{ik} = 0, \quad j \neq i \qquad (N6)$$

In terms of **A**, $b_{jk} = a_{jk}$, and $b_{ik} = A_{ik}$, which provides the relation:

$$\sum_k b_{jk} b_{ik} = \sum_k a_{jk} A_{ik}, \quad j \neq i \qquad (N7)$$

According to Theorem 1, the sum on the right is zero which is the required result.

## Appendix II: Sub-dimensional Cases

The ball set is the same as in Section I:

**B** = {$b_i$}, $b_i = (\mathbf{x_i}, r_i)$, $1 \leq i \leq d + 1$

Here, we give solutions to (1) for the cases $\det(\underline{V}) = 0$ with $\text{rank}(\underline{V}) = d - 1$, with $\underline{V}$ as defined above (P4). That is, the $d + 1$ position vectors of **B**, $\{\mathbf{x}_i\}$, which imply the d difference vectors, $\{\mathbf{x}_{ik}\}$, $1 \leq i \neq k \leq d$, do not span $\mathbf{R}^d$, but rather only $\mathbf{R}^{d-1}$. Under these conditions the one dimensional sub-space not spanned by $\{\mathbf{x}_{ik}\}$, can be represented by a single unit vector, $\hat{\mathbf{n}}$, that can be chosen such that $\hat{\mathbf{n}}.\mathbf{x}_{ik} = \mathbf{0}$ (see below S9).

Define the translated balls,

$$\mathbf{u}_i = \mathbf{x}_i - \mathbf{x}_{d+1}, \quad w_i = r_i - r_{d+1} \tag{S1a}$$

and similarly for the solution ball(s),

$$\mathbf{u} = \mathbf{x} - \mathbf{x}_{d+1}, \quad w = r + r_{d+1} \tag{S1b}$$

Although the treatment below is completely general, we suggest the convention that for all i, $w_i \geq 0$. Here, this amounts to assuming that $b_{d+1}$ has the smallest radius so that it is reduced to a point by the translation of (S1b) and the interpretation of the shifted solution radius, w, is not complicated by consideration of small negative w's. With or without this convention, the interpretation of the solution radius, r, is identical to that given in Section I.C so long as all $r_i \geq 0$, as discussed in Section I.A.

Also note that from (S1a), $\mathbf{u}_{d+1} = \mathbf{0}$, and $w_{d+1} = 0$. With these definitions, (1) becomes:

$$|\mathbf{u}_i - \mathbf{u}|^2 = (w_i + w)^2 \tag{S2}$$

and the, d, difference equations corresponding to (6) can be written:

$$\mathbf{u} \cdot \mathbf{u}_i + w w_i = (|\mathbf{u}_i|^2 - w_i^2)/2, \quad 1 \leq i \leq d \tag{S3a}$$

while for $i = d + 1$, we obtain the following directly from (S2) and (S1a):

$$|\mathbf{u}|^2 = w^2 \tag{S3b}$$

Any point in the $\mathbf{R}^{d-1}$ subspace spanned by $\{\mathbf{u}_i\} = \{\mathbf{x}_{ik}\}$ can be represented by some linear combination of the $\mathbf{u}_i$;

$$\mathbf{c} = \Sigma_j \, c_j \, \mathbf{u}_j, \quad 1 \leq j \leq d - 1 \tag{S4}$$

Note that any subset of d - 1 vectors from the set $\{\mathbf{u}_i\}$ is a sufficient basis, so in (S4) we omit $\mathbf{u}_d$. The solution vector(s) cannot be assumed to reside in the sub-space defined by $\{\mathbf{u}_i\}$ but can be expressed:

$$\mathbf{u} = \mathbf{c} + h\hat{\mathbf{n}} \tag{S5}$$

Substituting (S5) into (S3a) results in the following:

$$\Sigma_j \, c_j \mathbf{u}_i \cdot \mathbf{u}_j + w w_i = (|\mathbf{u}_i|^2 - w_i^2)/2 \tag{S6}$$

Note the absence of vector $\hat{\mathbf{n}}$, and scalar h, which have been annihilated due to the property $\hat{\mathbf{n}}.\mathbf{x}_{ik} = \hat{\mathbf{n}}.\mathbf{u}_i = \mathbf{0}$ (see below S9 ). With (S6) we have d equations with d variables to be determined: the d - 1 coefficients of c, and the scalar w. To find these, define the d x d matrix, $\underline{U}$, with elements $U_{ij} = \mathbf{u}_i \cdot \mathbf{u}_j$, $1 \leq i \leq d$, $1 \leq j \leq d - 1$ and $U_{id} = w_i$. Further define the d x 1 column matrices, $\underline{t}$, with elements, $t_i = (u_i^2 - w_i^2)/2$, and, $\underline{g}$, $g_i = c_i$, $1 \le i \le d - 1$, $g_d = w$, and the entire system can be expressed:

$$\underline{U}\underline{g} = \underline{t}, \text{ with formal solution, } \underline{g} = \underline{U}^{-1}\underline{t} \qquad (S7)$$

The coefficients, $c_j$, of $\mathbf{c}$, and scalar, w, are recovered from the solution column matrix $\underline{g}$.
To determine the scalar, h, note from (S5) and (S3b) we have, $w^2 = |\mathbf{c}|^2 + h^2$, or

$$h_\pm = \pm( w^2 - |\mathbf{c}|^2)^{1/2} \qquad (S8)$$

In terms of original variables:

$$\mathbf{x}_\pm = \mathbf{x}_{d+1} + \mathbf{c} + h_\pm \hat{\mathbf{n}} \qquad (S9a)$$
$$r = w - r_{d+1} \qquad (S9b)$$

Unlike the full dimensional cases of Sections I and II, there is only one solution radius, but still two solution vectors. The solutions therefore continue to come in pairs, which can be imaginary if the discriminate of (S8) is negative. Since there is only one solution radius, real solutions are either both "large negative" or neither. If all of the $w_i$ are zero (resulting from all $r_i$ equal), matrix $\underline{U}$ will be singular and there are no solutions. For 3-d and higher, matrix $\underline{U}$ can be singular even if the $w_i$ are not all equal. Such cases are beyond the scope of the present work, but solutions are still possible.

To use (S7-9) a method of finding vector, $\hat{\mathbf{n}}$, is still required. In 2-d, we have only one linearly dependent difference vector, say, $\mathbf{u}_1 = u_{11}\hat{\mathbf{e}}_1 + u_{12}\hat{\mathbf{e}}_2$. If either component of $\mathbf{u}_1$ is zero, the normal vector is along that direction, and in general, $\mathbf{n} = u_{12}\hat{\mathbf{e}}_1 - u_{11}\hat{\mathbf{e}}_2$. For 3-d, we have two, say $\mathbf{u}_1$ and $\mathbf{u}_2$, and $\mathbf{n} = \mathbf{u}_1 \wedge \mathbf{u}_2$. Note that in both cases, the condition, $\mathbf{n} \cdot \mathbf{u}_i = 0$, is fulfilled for $1 \le i \le d$. In higher dimensions one can use the method described in Appendix I, except here we need a normal vector in $\mathbf{R}^d$, so we use matrix $\underline{V}$ of Section II, but with the bottom row nulled. The components of $\mathbf{n}$ are then the cofactors of the bottom row. In all cases, we define $\hat{\mathbf{n}} = \mathbf{n}/|\mathbf{n}|$ for use above.